\documentclass[aps,pra,a4paper,showpacs,twocolumn,floatfix]{revtex4}
\usepackage[latin1]{inputenc}
\usepackage{amsfonts}
\usepackage{amssymb}
\usepackage{amsmath}
\usepackage{calc}
\usepackage{graphicx}
\usepackage{epsfig}
\usepackage{bm}
\usepackage{ulem}
\usepackage{setspace}
\usepackage{subfigure}
\usepackage{epstopdf}

\begin{document}

\title{Quantum Repeaters based on Single Trapped Ions}
\date{\today}
\author{Nicolas Sangouard$^{1}$, Romain Dubessy$^{1}$, and Christoph Simon$^{2}$}
\affiliation{%
$^{1}$Laboratoire Matériaux et Phénomènes Quantiques CNRS, UMR7162, Université Paris Diderot, France\\
$^{2}$Group of Applied Physics, University of Geneva, Switzerland
}

\begin{abstract}
We analyze the performance of a quantum repeater protocol
based on single trapped ions. At each node, single trapped
ions embedded into high finesse cavities emit single
photons whose polarization is entangled with the ion state.
A specific detection of two photons at a central station
located half-way between two nodes heralds the entanglement
of two remote ions. Entanglement can be extended to long
distances by applying successive entanglement swapping
operations based on two-ion gate operations that have
already been demonstrated experimentally with high
precision. Our calculation shows that the distribution rate
of entanglement achievable with such an ion-based quantum
repeater protocol is higher by orders of magnitude than the
rates that are achievable with the best known schemes based
on atomic ensemble memories and linear optics. The main
reason is that for trapped ions the entanglement swapping
operations are performed deterministically, in contrast to
success probabilities below 50 percent per swapping with
linear optics. The scheme requires efficient collection of
the emitted photons, which can be achieved with cavities,
and efficient conversion of their wavelength, which can be
done via stimulated parametric down-conversion. We also
suggest how to realize temporal multiplexing, which offers
additional significant speed-ups in entanglement
distribution, with trapped ions.
\end{abstract}

\maketitle

\section{Introduction}
The distribution of entanglement over long distances is
difficult because of unavoidable transmission losses and
the no-cloning theorem for quantum states. One possible
solution is the use of quantum repeaters \cite{Briegel98},
which are based on the heralded creation and storage of
entanglement for elementary links of moderate length,
followed by entanglement swapping operations that allow one
to extend the distance of entanglement.

The most widely known approach to quantum repeaters
combines quantum memories based on atomic ensembles and
entanglement swapping operations using linear optics.
Building on the initial proposal of Ref. \cite{Duan01},
there has been a large amount of experimental \cite{atexp}
and theoretical \cite{Jiang07,attheo,Sangouard08} work
towards realizing long-distance entanglement distribution
in this way. This approach is attractive because it uses
relatively simple elements. However, as quantum technology
progresses, it is natural to also consider other possible
physical systems. For example, there have been recent
proposals for the realization of quantum repeaters with NV
centers in diamond \cite{Childress06} and with spins in
quantum dots \cite{SimonNiquet07,VanLoock08}.

Trapped ions were one of the first systems to be proposed
for quantum computation \cite{Cirac95}. Since then, many
experiments have been realized that demonstrated key
ingredients for quantum computing, including the
realization of quantum logic gates with increasing
precision \cite{Monroe95, SchmidtKaler03, Leibfried03,
Benhelm08}, the creation of multi-qubit entanglement
\cite{Turchette98, Roos04,Leibfried05}, the implementation
of the Deutsch-Jozsa algorithm \cite{Gulde03}, the
realization of entanglement purification
\cite{Chiaverini04, Reichle06}, the accomplishment of
atomic-qubit quantum teleportation \cite{Riebe04, Barrett04},
the realization of deterministic entanglement swapping \cite{Riebe08}
as well as the demonstration of very high-efficiency detections
with \cite{Hume07} and without \cite{Myerson08} ancilla qubit.
In addition, ion-photon entanglement has been created
\cite{Blinov04} and subsequently used to entangle distant
ions \cite{Moehring07}.

Motivated by this impressive body of work, we here analyze
in detail the achievable performance of quantum repeaters
based on trapped ions. We show that the distribution rates
of entanglement offered by repeaters based on single ions
are significantly superior compared to the ones achieved
with atomic ensemble based schemes. The main reason is that
entanglement swapping operations can be performed
deterministically for trapped ions. In contrast, the
success probability for entanglement swapping is bounded by
1/2 for schemes using Bell measurements based on linear
optics \cite{Calsamiglia01}.

The repeater protocol that we envision requires both an
efficient collection of the emitted photons and an
efficient conversion of their wavelength to the telecom
wavelength around 1.5 $\mu$m where the losses in optical
fibers are at their minimum. In order to improve the
collection of photons emitted by a single ion, one can couple
this ion to a high-finesse cavity. Individual ions have been
coupled to high-finesse cavities experimentally
\cite{Mundt02, Keller03} and theoretical proposals \cite{Maurer04}
have been realized to make very efficient the photon emission
probability into the cavity mode using realistic cavity parameters,
cf. below. The frequency conversion might be realized using
stimulated parametric down-conversion. This is in fact, the inverse
process of the coherent up-conversion that was demonstrated
for single photons in Ref. \cite{Tanzilli05}, cf. below.

The performance of atomic ensemble based quantum repeaters
can be greatly enhanced by temporal multiplexing
\cite{Simon07} using multi-mode memories \cite{Afzelius08}.
We will suggest how to implement analogous temporal
multiplexing for trapped ions using ion transport methods
that have been developed in the context of quantum
computing.

This paper is organized as follows. In the next section, we
present the achievable distribution rates for a repeater
protocol based on single trapped ions and we compare them
to the ones achievable with atomic ensembles. The third
section is devoted to implementation issues. In the fourth
section we present an approach to implement temporal
multiplexing. The fifth section contains our conclusions.

\section{Efficiency of repeaters with trapped ions}

\begin{figure}
{\includegraphics[scale=0.18]{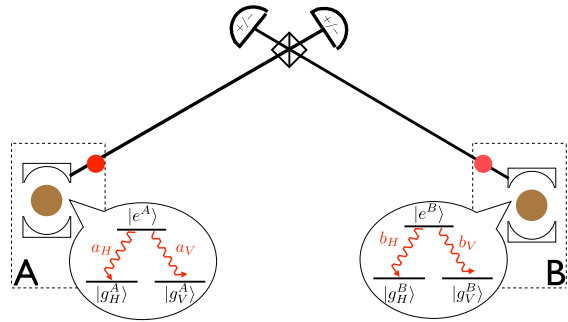} \caption{(Color
online) Setup for entanglement creation based on two-photon
detection of remote ions (brown dots) embedded into
cavities. Each ion emits a photon, whose polarization is
entangled with the atomic state, leading to the state
$|\Psi^A\rangle\otimes |\Psi^B\rangle$ of Eq. (1). The two
photons, one coming from location $A,$ the other one from
$B$ are combined on a polarizing beam splitter to be
further detected in a polarization basis rotated by 45
degrees with respect to the $H-V$ basis. The coincident
detection of two photons in the corresponding modes $d_+$
and $\tilde{d}_+$, for example, projects the two ions into
the entangled state $|\psi^{AB}_+\rangle$ of Eq.
(2).}\label{fig1}}
\end{figure}

Let us recall how two remote ions at locations A and B can
be entangled via the detection of two photons as
proposed in \cite{Feng03, Duan03, Simon03}. Note that two
remote ions can also be entangled based on the single-photon
detection \cite{Cabrillo99}. For a discussion of
the advantages and disadvantages of schemes based
on two-photon detections versus schemes based on single-photon
detections, see e.g. Refs. \cite{Zippilli08, Sangouard08}. Each ion is described by a
lambda system of three states, as shown in Fig. \ref{fig1}.
From the excited state $|e^A\rangle$ ($|e^B\rangle$) the
ion located at A (B) can decay into two degenerate
metastable states, say the states $|g^A_V\rangle$ and
$|g^A_H\rangle$ ($|g^B_V\rangle$ and $|g^B_H\rangle$) by
emitting a photon with a well defined polarization, say
either vertical corresponding to the mode $a_V$ or
horizontal corresponding to $a_H$ ($b_V$ and $b_H$
respectively). The A and B ions are both excited
simultaneously, such that the emission of a photon by each
ion leads to the state
\begin{eqnarray}
&&|\Psi^A\rangle\otimes |\Psi^B\rangle = \\
&&\nonumber \frac{1}{2}\left(|g_H^A\rangle a^{\dagger}_H+|g^A_V\rangle a^{\dagger}_V\right) \otimes \left(|g_H^B\rangle b^{\dagger}_H+|g^B_V\rangle b^{\dagger}_V\right)|0\rangle
\end{eqnarray}
with $|0\rangle$ the vacuum state. A probabilistic Bell
state analysis can be performed by combining the two
emitted photons on a  polarizing beam spitter (PBS) at a
central station located half-way between $A$ and $B$ and by
counting the photon number in each output modes $d_\pm =
\frac{1}{\sqrt{2}}(a_H \pm b_V),$ $\tilde{d}_\pm =
\frac{1}{\sqrt{2}}(b_H\pm a_V)$. Such Bell analysis
projects non-destructively the two ions into an entangled
state. For example, the detection of two photons, one in
each modes $d_+$ $\tilde{d}_+$,  leads to the entangled
state
\begin{equation}
\label{eq2}
|\psi^{AB}_+\rangle=\frac{1}{\sqrt{2}}\left(|g_H^Ag_H^B\rangle +|g^A_Vg^B_V\rangle\right).
\end{equation}
In the ideal case, the probability for such an event is
$1/8.$ Taking into account the coincidences between
$d_-$-$\tilde{d}_+,$ $d_+$-$\tilde{d}_-$ and
$d_-$-$\tilde{d}_-$ combined with the appropriate one-qubit
operations, the probability to create the state (\ref{eq2})
is $1/2$. This way of creating entanglement was
demonstrated experimentally in Ref. \cite{Moehring07}. Note
that the photon collection efficiency was quite low in
these experiments, which did not have cavities around the
ions.

We now calculate the time needed for entanglement creation.
Let us denote by $p$ the success probability for an ion to
emit a photon, which includes the probability to prepare
the ion in the excited state, the spontaneous emission of a
photon into the cavity mode and coupling into the fiber, as
well as the frequency conversion to match the telecom
wavelength. The probability to get the expected twofold
coincidence is thus given by $P_0=\frac{1}{2}p^2 \eta_t^2
\eta_d^2$ where $\eta_t=e^{-L_0/(2 L_{att})}$ is the fiber
transmission with the attenuation length $L_{att}$ (we use
$L_{att}= 22$ km, corresponding to losses of 0.2 dB/km,
which are currently achievable at a wavelength of 1.5
$\mu$m) and $\eta_d$ is the detection efficiency.
Entanglement creation attempts can be repeated at time
intervals given by the communication time $L_0/c$, cf. Ref.
\cite{Simon07}. As a consequence, the average time required
to entangle two ions separated by a distance $L_0,$ is
given by
\begin{equation}
\label{eq3} T_{link}=\frac{L_0}{c}\frac{1}{P_0}.
\end{equation}
Here $c = 2 \times 10^8$ m/s is the photon velocity in the fiber.\\

\begin{figure}
{\includegraphics[scale=0.3]{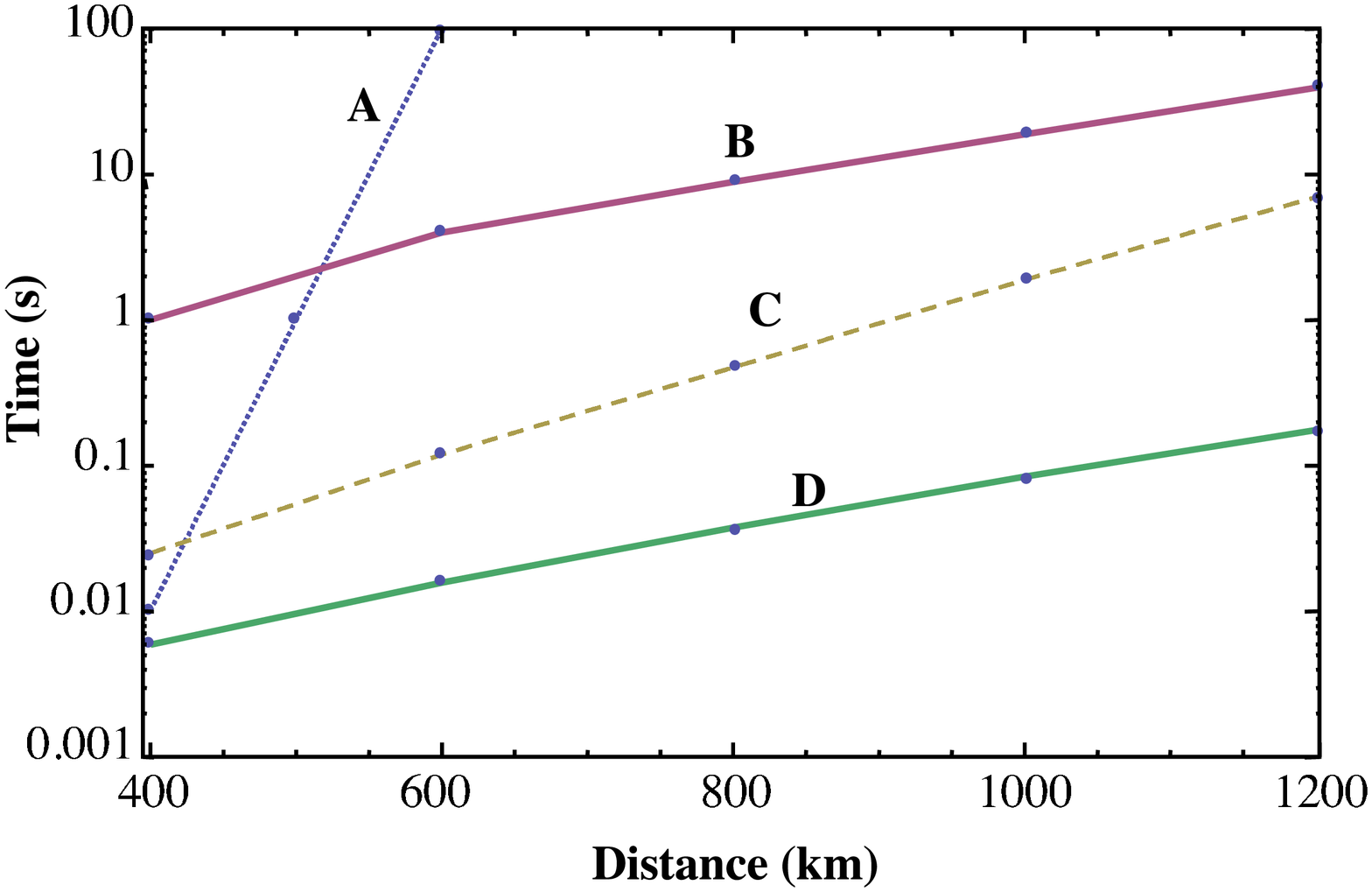} \caption{(Color
online) Performance of quantum repeaters based on single
ions versus atomic ensembles. The quantity shown is the
average time for the distribution of one entangled pair for
the given distance. Curve A: as a reference, the time
required using direct transmission of photons through
optical fibers, with losses of 0.2 dB/km, corresponding to
the best available telecom fibers at a wavelength of 1.5
$\mu$m, and a pair generation rate of 10 GHz. Curve B:
protocol based on atomic ensembles of Ref.
\cite{Sangouard08}. High-fidelity entangled pairs are
generated locally, and entanglement generation and swapping
operations are based on two-photon detections. We have
assumed memory and detector efficiencies of 90\%. We
imposed a maximum number of 16 links in the repeater chain
(see \cite{Sangouard08} for details). This approach leads
to a repeater protocol that, as far as we know, achieves
the highest entanglement distribution rate with atomic
ensembles and linear optics. Curve C and D: protocol based
on single ions with 8 and 16 links respectively. We have
assumed a success probability for the ion to emit a photon
of $p=90\%$ requiring high-finesse cavity, cf. text.}
\label{fig2}}
\end{figure}

The entanglement can further be distributed over longer
distances by using successive entanglement swapping
operations between elementary links. Such swapping
operations require a local Bell state analysis, applied
e.g. on the two ions located at B to entangle the ions
located at A and C. Bell states have recently been prepared
deterministically from the computational basis with a
very-high fidelity \cite{Benhelm08}. Applied on the four
Bell states, this protocol transforms each of them into a
product state in the computational basis. The measurement
of the individual ion states then leads to the desired Bell
analysis. The success probability for entanglement swapping
reduces in this case to the detection efficiency of ions,
which is essentially equal to one. The time for the
swapping and detection can realistically be much shorter
than the time required for entanglement creation
(\ref{eq3}), cf. below, such that the total time for the
distribution of an entangled pair over the distance $2 L_0$
is given by
\begin{equation}
\label{eq4}
T_{2L_0} \approx \frac{3}{2} \frac{L_0}{c} \frac{1}{P_0}= \frac{3L_0}{c} \frac{1}{p^2\eta_t^2\eta_d^2}.
\end{equation}
The factor 3/2 arises because entanglement has to be
generated for two links before the entanglement connection
can be performed. If the average waiting time for
entanglement generation for one link is $T$ , there will be
a success for one of the two after $T /2$ ; then one still
has to wait a time $T$ on average for the second one,
giving a total of $3T /2$. This simple argument gives
exactly the correct result in the limit of small $P_0$
\cite{Collins07,Brask08}. For a quantum repeater with $n$
nesting levels, analogous factors arise at each level. They
are no longer exactly equal to 3/2 in the general case
because the waiting time distribution for establishing an
individual higher-level link is no longer simply
exponential, but numerical results show that this remains a
good approximation \cite{Jiang07,Brask08}. (Note that the
factors certainly all lie between 1 and 2.) The average
time for the distribution of an entangled pair over the
distance $L=2^n L_0$ is then approximately given by
\begin{equation}
\label{Ttot}
T_{tot} \approx \left(\frac{3}{2}\right)^n \frac{L_0}{c} \frac{1}{P_0}=\frac{3^n}{2^{n-1}} \frac{L_0}{c} \frac{1}{p^2\eta_t^2\eta_d^2}.
\end{equation}

The performance of such a quantum repeater based on single
ions is shown in Fig. \ref{fig2}. In the same figure we
also show the performance of the best atomic ensemble based
protocol known to us \cite{Sangouard08}. In this approach,
one first locally generates high-fidelity entangled pairs
of atomic excitations that are stored in nearby ensembles.
Then long-distance entanglement is generated and swapped
via two-photon detections. As in Ref. \cite{Sangouard08},
we have limited the maximum number of links used to 16 for
all protocols, to have link numbers for which it is
plausible that entanglement purification may not be
necessary. Note however that entanglement purification
has already been implemented for trapped ions
\cite{Chiaverini04,Reichle06}.

\begin{figure}
{\includegraphics[scale=0.30]{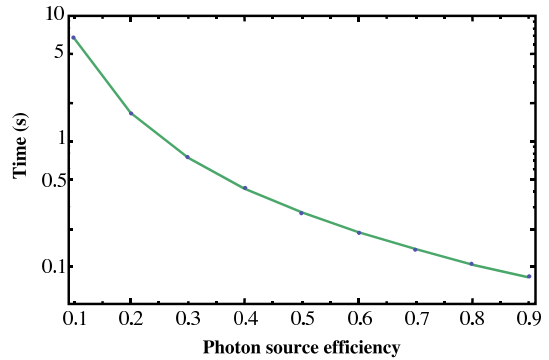} \caption{(Color
online) Robustness of a repeater based on single trapped
ions with respect to the success probability for an ion to emit
a photon $p$ (photon source efficiency) which includes the
probability to prepare the ion into the cavity mode and coupling
into the fiber, as well as the frequency conversion to match the
telecom wavelength. The quantity shown is the average time
for the distribution of an entangled pair over 1000km for a repeater
with 16 elementary links (Eq. (\ref{Ttot})).}\label{fig3}}
\end{figure}

Fig. 2 shows that the distribution rate that can be
achieved with single ions is higher by orders of magnitude
than the one obtained with atomic ensembles. As mentioned
before, the most important factor explaining this
improvement is that entanglement swapping operations are
performed deterministically for the ions, whereas each
swapping operation is performed at most with a probability
1/2 using linear optical elements. Another reason is that
the state generated locally with atomic ensembles, which
should ideally be a state of two maximally entangled atomic
excitations, in fact possesses no-excitation and
single-excitation components. Even if these undesired
components can be reduced by partial memory read-out
\cite{Sangouard08}, they still limit the achievable
distribution rate of entanglement.

In Fig. 2 we have assumed that the ions are very efficient
sources of single photons ($p=90\%$), in order to keep the
assumptions comparable with the ones made for the
atomic-ensemble based scheme in Ref. \cite{Sangouard08},
where the memory efficiency was taken to be $\eta_m=90\%$.
However, it should be pointed out that the average time for
the distribution of an entangled pair (see Eq.
(\ref{Ttot})) scales only like $\frac{1}{p^2}$, such that
even with $p=30\%,$ one needs less than $T_{tot}=740$ ms to
distribute an entangled pair over 1000 km using 16 links,
which is still shorter than the time achievable with atomic
ensembles. Fig. \ref{fig3} gives the average time required to
distribute an entangled pair for various values of $p$.
Note that atomic ensemble based schemes are much
more sensitive to a reduction in $\eta_m$, because it
intervenes in every swapping operation.

\section{Implementation}

To achieve a high efficiency of photon collection, one can
embed the ion within a cavity. The spontaneous emission
emitted into the cavity mode is enhanced by the Purcell
factor
\begin{equation}
F_P=\frac{3\ell \lambda^2}{2\pi^2V_0}\mathcal{F}
\end{equation}
with $\mathcal{F}$ the finesse of the cavity, $\ell$ its
length, $\lambda$ the free-space wavelength and $V_0$ the
mode volume of the cavity (which is of order
$\ell^2\lambda$ for a confocal cavity with a waist of order
$\sqrt{\ell\lambda}$). The collection efficiency
$\frac{F_P-1}{F_P}$ can then be made as large as desired
for large enough Purcell factor $F_P$, which requires a
high finesse $\mathcal{F}$ and small mode volume $V_0$.
Note that a Purcell factor of 2 was already achieved
experimentally for a trapped ion in a cavity in Ref.
\cite{Mundt02}.

\begin{figure}
{\includegraphics[scale=0.19]{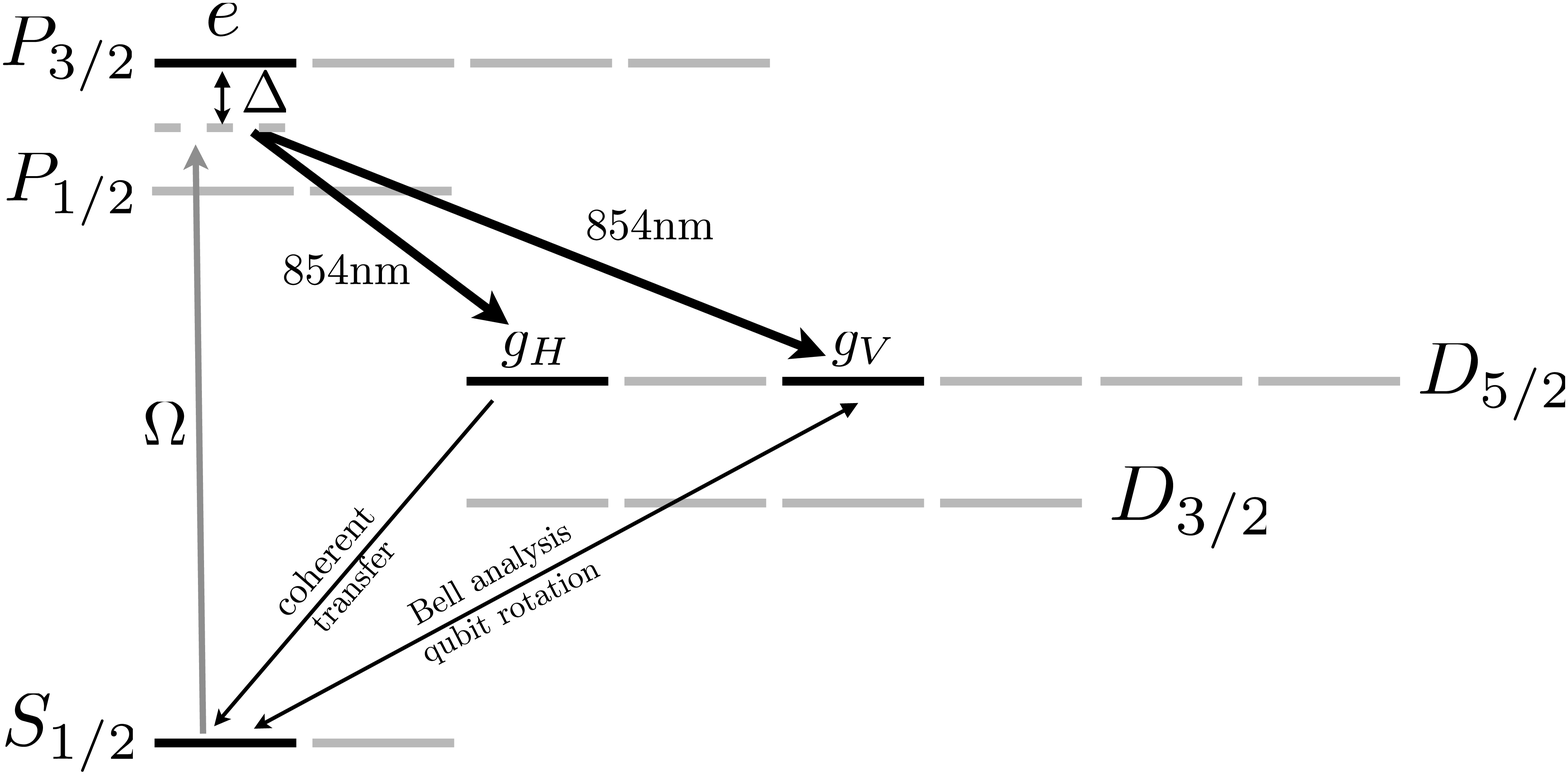}
\caption{Relevant levels for $^{40}$Ca$^+$.}
\label{fig4}}
\end{figure}

For concreteness, we focus on the realization of the
studied repeater protocol with $^{40}$Ca$^+$ ions (the
relevant states are presented in Fig. \ref{fig4}) even if
other species should not be excluded. Following the
proposal of Ref. \cite{Simon03}, one could prepare the ions
in one of the $P_{3/2}$ sublevels to serve as excited state
$|e\rangle.$ For $|g_H\rangle$ and $|g_V\rangle,$ one could
use two sublevels of $D_{5/2}$ which are coupled to
$|e\rangle$ by orthogonally polarized photons at 854 nm.
Ideally, the coupling strengths for the transitions
$|e\rangle$-$|g_H\rangle$ and $|e\rangle$-$|g_V\rangle$
should be equal and the cavities should be designed such that
the two polarizations are equally supported otherwise the
probability to create the state (\ref{eq2}) is reduced. Note that in principle,
if the coupling strengths are not equal, they might be compensated
by appropriate cavity couplings. The characteristic lifetime of
the sublevels of $D_{5/2}$ is up to 1 s, which is compatible
with the average time required for the distribution of an entangled
pair for all the distances considered in Fig. \ref{fig2} for repeaters
with 16 links (see curve D). If longer memory times are
required, e.g. for repeaters with 8 links (see curve C),
one could coherently transfer these states to the sublevels
associated to $S_{1/2},$ cf. below.

The state $P_{3/2}$ decays preferentially to $S_{1/2}$,
which at first sight seems to limit the achievable photon
collection efficiency. To overcome this limitation, it has
been proposed in Ref. \cite{Maurer04} to couple the ground
state $S_{1/2}$ directly to sublevels of $D_{5/2}$ through
a Raman process by choosing a pump laser far detuned from
the $S_{1/2}$-$P_{3/2}$ transition, i.e. $\Delta \gg
\Omega$ with $\Delta$ the detuning and $\Omega$ the pump
Rabi frequency. It is shown in Ref. \cite{Maurer04} that
with this approach one can achieve a photon emission
probability into the cavity mode of 95\% for a realistic
cavity. The achievable photon repetition rate of 20 kHz
proposed in Ref. \cite{Maurer04} is higher than
$\frac{c}{L_0}$ as soon as the elementary links are longer
than $L_0=15$ km. As a consequence, for the considered
distances and link numbers ($L_0 \geqslant 25$ km), the
average time for entanglement creation is limited by the
communication time, which is in agreement with Eq.
(\ref{eq3}).

We have assumed that the wavelength of the photons emitted
by the ions is converted to a telecom wavelength around 1.5
$\mu$m, in order to profit from the optimal transmission of
optical fibers in that range. Frequency conversion at the
single photon level was already demonstrated in Ref.
\cite{Tanzilli05} with an intrinsic efficiency of $56 \%$.
In this experiment the frequency of the photons was
up-converted in order to achieve a better detection
efficiency. However, the inverse process, which is
parametric down-conversion with a single-photon pump, but a
strong laser stimulating emission into one of the two
down-converted modes, can be performed with the same
efficiency (due to unitarity). It should be possible to
bring the conversion efficiency close to one using stronger
non-linearities and a stronger stimulation laser, and of
course minimizing all optical and coupling losses. Note
that this conversion process preserves entanglement, as was
already demonstrated in Ref. \cite{Tanzilli05}.

A Bell state analysis is required for the entanglement
swapping operations. Following the proposal of Ref.
\cite{Sorensen99}, two $^{40}$Ca$^+$ ions have recently
been prepared deterministically in a Bell state (the two
qubit states are sublevels of $S_{1/2}$ and $D_{5/2}$) with
a fidelity greater than 99\% on a time scale of the order
of 50 $\mu$s \cite{Benhelm08}. The two ions are placed
close to each other such that they interact through the
Coulomb interaction giving rise to a common spatial
vibration. A collective irradiation with the appropriate
bichromatic field allows one to prepare deterministically
the desired Bell state from a given state of the
computational basis \cite{Sorensen99}. Such an experiment
could be used to perform the required Bell state analysis
in the following manner. The two ions located at each node
could be embedded within the same cavity. The distance between
them has to be small enough such that they interact efficiently through Coulomb
interaction but large enough to allow one an individual addressing
of each of them with laser beams. Such an addressing is essential
for entanglement creation, i.e. for the targeted emission of a photon by
one of the two ions. An optical switch could be used to send the
emitted photon to the desired central station. A typical distance of
$\sim$ 8 $\mu$m separating the two ions \cite{Benhelm08} with laser
beams focused to $\sim$ 2 $\mu$m might be well suited. For
entanglement swapping, one could first transfer coherently the
population of $|g_H\rangle$ to a sublevel of
$S_{1/2},$ as in Refs. \cite{Schmidt03, Roos99} requiring a
time scale of $\sim$10 $\mu$s. We then use the appropriate
bichromatic field on the transition $S_{1/2}$-$|g_V\rangle$ as in
Ref. \cite{Benhelm08} such that each Bell state will be
transformed into a given state of the computational basis.
This takes $\sim$ 50 $\mu$s. We finally measure the state
of each ion independently. This detection could be done by
measuring resonance fluorescence from the auxiliary state
$P_{1/2}$ that is strongly coupled to $S_{1/2}$ with a
laser field at 397 nm and decays back only to that same
state \cite{SchmidtKaler03}. Such measurement has been
preformed recently \cite{Myerson08} in the same system and it
takes in average 145$\mu$s with a photon collection of 0.2\%.
Such characteristic time can realistically be reduced to a few tens of
$\mu$s by optimizing the collection efficiency \cite{Simon03}.
All in all, an entanglement swapping operation should be much
shorter than the average time for the entanglement creation
($T_{link}>1$ ms for $L \geqslant 400$km) justifying the
formulas (\ref{eq4})-(\ref{Ttot}).

To exploit the entanglement, it is essential to be able to
detect the states of the ions in different bases (e.g. for
a Bell test or for quantum key distribution). The necessary
rotations could be performed by first coherently
transferring $|g_H\rangle$ to the $S_{1/2}$ sublevels as
before and then applying the appropriate pulses on the
transition involving that state and $|g_V\rangle$. As said
before, these transformations can be performed in $\sim$ 10
$\mu$s.

\section{additional speed-up via temporal multiplexing}

\begin{figure}
{\includegraphics[scale=0.3]{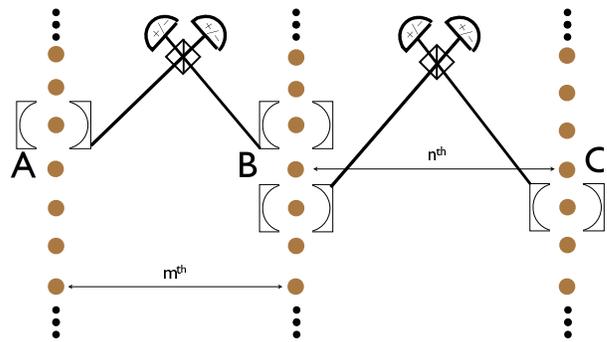} \caption{(Color
online) Setup for temporal multiplexing. Chains of ions are
transported through cavities such that the ions are excited
one by one when they interact with the cavity mode. The
ions of the chain B are alternatively excited in the upper
cavity for entanglement creation between A and B or in the
lower cavity for entanglement creation between B and C. If
entanglement has been established between the $m$$^{th}$
ions for the link A-B and between the $n$$^{th}$ ions for
B-C, entanglement swapping is done by performing a Bell
state analysis on the $m$$^{th}$ and $n$$^{th}$ ions of the
B chain. }\label{fig5}}
\end{figure}

As seen before, the creation of entanglement between
neighboring nodes A and B is conditioned on the outcome of
photon detections at a station located half-way between the
nodes. To profit from a nested repeater, the entanglement
swapping operations can only be performed once one knows
the relevant measurement outcomes. This requires a
communication time of order $L_0/c.$ If one can perform a
number $N$ of entanglement creation attempts per elementary
link within the time interval $L_0/c,$ one can decrease
the average time for entanglement creation $T_{link}$ by a
factor of order $N$. Such temporal multiplexing has
initially been proposed for quantum repeaters based on
atomic ensembles \cite{Simon07}, and since then a
particularly efficient quantum storage protocol has been
developed \cite{Afzelius08} for this purpose. We here
propose a realization of the same basic idea for quantum
repeaters based on single ions.

Consider two links, say A-B and B-C, allowing one to
connect the A and C nodes by entanglement swapping, see
Fig. \ref{fig5}. At each location A, B and C a chain of
ions within a segmented trap can be moved through a cavity
by applying appropriate control electric fields to the
various segments \cite{Barrett04, Rowe02, Huber08} such
that the internal state of the ions is preserved. Further
suppose that the distance between two successive ions is
larger than the waist radius of the cavity mode such that
one can selectively excite each ion when it interacts with
the cavity mode in order to force it to emit a photon.

The ions located at B are used as sources for entanglement
creation between both the A-B and B-C links in the
following way. Suppose that the chains located at A and C
are composed of $N$ ions. The chain B possesses $2N$ ions
which are excited alternatively in the upper cavity and in
the lower one for entanglement creation between A-B and
between B-C locations respectively. If there are two
detections behind the central PBS located between A and B
for the $m$$^{th}$ ions for example, then we know that
these ions are entangled. Running the same protocol for
another pair of ions, there may be similar detections
between B and C locations associated to the $n$$^{th}$
ions. One then performs entanglement swapping by applying
the appropriate operations on the $m$$^{th}$ and $n$$^{th}$
ions of the B chain. This can be done by addressing
individually the $m$$^{th}$ and the $n$$^{th}$ ions with
the appropriate bichromatic field \cite{Sorensen99}, thus
realizing the Bell state analysis described in the previous
section.

Single $^{40}$Ca$^+$ ions have already been transported
from a loading zone to a cavity interaction region
separated by more than 20 mm in a characteristic time of 4
ms \cite{Keller03}. This was realized using a segmented
trap composed of 5 pairs of electrodes by successively
ramping the electrode voltages. Faster transports were
realized along $\sim$1 mm with a characteristic time of 50
$\mu$s without loss of coherence and with negligible
excitation of the ion's motion \cite{Rowe02, Barrett04,
Huber08}. The number of attempts that can be achieved per
time interval $L_0/c$ is thus likely to be limited by the
characteristic time of the Raman process, rather than by
the speed of ion transport. For example, the 20 mm long
cavity considered in Ref. \cite{Maurer04}, which is
compatible with the characteristics of the trap reported in
Ref. \cite{Keller03}, gives a photon repetition rate of 20
kHz, which would allow 10 attempts per time interval
$L_0/c$ for 1000 km and 8 links. This would increase the
entanglement distribution rate by the same factor of 10.
For higher repetition rates, one needs to decrease the
duration of the Raman process $\tau$, which has to fulfill
$\frac{\Omega g}{\Delta} \tau \sim \pi$ to insure an
efficient population transfer. This can be done by
increasing the $g$ factor, i.e. by decreasing the cavity
length $l_c.$ (The ratio $\Omega/\Delta$  has to be kept
smaller than 1 to guarantee that no population will be
transferred to the excited state). Considering e.g. a
cavity length of $l_c=6$ mm as described in Ref.
\cite{Keller03}, the achievable distribution rate increases
by a factor 30. If one chooses $l_c=1$ mm, which might
still be compatible with microtrap dimensions
\cite{microtrap}, one gets an improvement of the rate by a
factor of 200.

\section{Conclusion}

We have shown that trapped ions are very promising systems
for the implementation of quantum repeaters. In fact, the
achievable performance for a relatively basic trapped ion
quantum repeater protocol greatly exceeds the best atomic
ensemble based protocol known to us. This is mostly due to
the fact that a deterministic Bell state analysis can be
performed for trapped ions using current technology. We
have argued that this performance could further be improved
very significantly using temporal multiplexing based on ion
transport techniques that have been developed with quantum
computing applications in mind. The requirements for
implementing practically useful quantum repeaters, while
technologically challenging, are much more modest than for
the realization of fault-tolerant quantum computation. We
suggest that this is an interesting intermediate goal that
the ion trapping community should keep in mind.

\begin{acknowledgments}
We thank T. Coudreau, N. Gisin, L. Guidoni, and D. Lucas
for helpful discussions. This work was supported by the EU
Integrated Project {\it Qubit Applications}, the Swiss NCCR
{\it Quantum Photonics} and the French National Research
Agency (ANR) project ANR-JC05\_61454.
\end{acknowledgments}


\end{document}